\documentclass[10pt, aps, twocolumn]{revtex4}
\usepackage{graphicx}
\usepackage{amsmath}

\begin{document}

\title{A model of magnetic friction with the infinite-range interaction }

\author{Hisato \surname{Komatsu}}
\email[Email address: ]{komatsu.hisato@nims.go.jp}

\affiliation{Research Center for Advanced Measurement and Characterization, National Institute for Materials Science, Tsukuba, Ibaraki 305-0047, Japan}

\begin{abstract}

We investigate a model of magnetic friction with the infinite-range interaction by mean field analysis and a numerical simulation, and compare its behavior with that of the short-range model that we considered previously [H.~Komatsu, Phys.\ Rev.\ E.\ \textbf{100}, 052130 (2019)]. This infinite-range model always obeys the Stokes law when the temperature is higher than the critical value, $T_c$, whereas it shows a crossover or transition from the Dieterich--Ruina law to the Stokes law when the temperature is lower than $T_c$. 
Considering that the short-range model in our previous study shows a crossover or transition irrespective of whether the temperature is above or below the equilibrium transition temperature, the behavior in the high-temperature state is the major difference between these two models.

\end{abstract}

\maketitle

\section{Introduction \label{Introduction} }

Friction is an important subject in solid-state and applied physics~\cite{BC06, KHKBC12}. 
The Amontons--Coulomb law, in which frictional force $F$ is independent of relative velocity $v$, has long been used for the friction between solid surfaces. However, Coulomb himself pointed out that actual materials violate this law slightly~\cite{PP15}. This violation was studied several decades ago, and the empirical modification of the Amontons--Coulomb law known as the Dieterich--Ruina law~\cite{BC06, KHKBC12, Ruina83, Dieterich87, DK94, HBPCC94, Scholz98} was established.
 In the steady state, this law is expressed as 
\begin{equation}
F = A \log v + B,
\label{DRlaw}
\end{equation}
where $A$ and $B$ are constants. The term $A \log v$ is the difference from the Amontons--Coulomb law.

Despite these phenomenological or empirical studies, the microscopic mechanism of the friction is not fully understood, and various factors affecting friction, such as lattice vibration and the motion of electrons, have been considered~\cite{MDK94, DAK98, MK06, PBFMBMV10, KGGMRM94}. In particular, magnetic friction, which is the frictional force caused by the magnetic interaction between spin variables, has attracted much interest~\cite{WYKHBW12, CWLSJ16, LG18}, and several types of statistical mechanical model of this phenomenon have been proposed~\cite{KHW08, Hucht09, AHW12, HA12, IPT11, Hilhorst11, LP16, Sugimoto19, FWN08, DD10, MBWN09, MBWN11, MAHW11}.

In these models, the important behaviors of the system, such as the relation between the frictional force and the relative surface velocity, differ. Depending on the model, the system may obey the Amontons--Coulomb law~\cite{KHW08, Hucht09, AHW12,HA12} or the Stokes law~\cite{MBWN09,MBWN11}, while in some models the relation shows a crossover between these two laws~\cite{MAHW11}. In our previous model, the relation shows a crossover or transition from the Dieterich--Ruina law to the Stokes law~\cite{Komatsu19}. To compare these studies with the friction of actual magnetic materials or typical solid surfaces, the reason this difference appears should be considered first. However, these models differ from each other in many ways, such as the form of spin variables, the definition of the dynamics, and the interaction range. Hence, we need to investigate how each factor affects the friction.   

In this study, we introduce a model in which the spins interact with each other by the infinite-range interaction. This model resembles the short-range model in our previous study, except for the interaction range. Namely, we use Ising spin variables, and the system creates an ``antiferromagnetic order,'' which prevents the relative motion of the lattices as if the magnetic order is the potential barrier, and the dynamics are defined by mixing the Metropolis method and the Langevin equation. Comparing this model and our previous model, we investigate how the interaction range affects the magnetic friction.
In the realistic magnetic systems, the dipolar interaction is more important example of the long-range interaction. However, considering the relation between this interaction and the magnetic friction is thought to be difficult because it makes complicated magnetic structures depending on the condition\cite{DMW00,RRT07,MM10}. Furthermore, general long-range interaction systems need $O(N^2)$ computational complexity, so the numerical simulation of large size system itself is difficult. Indeed, previous researches on the magnetic friction caused by the dipolar interaction are limited to the relatively simple systems, such as the system with one tip and one chain\cite{MBWN09}.
 In the case of the infinite-range interaction system, on the other hand, consideration is easier because the computational complexity of the numerical simulation is $O(N)$, and the behavior at the thermodynamic limit can be investigated by mean field analysis. Moreover, as we will explain in the next section, we can construct the infinite-range interaction model which construct the magnetic structure similar to the antiferromagnetic order of the short-range model. This point enables us to consider the effect of the interaction range, without concern about the difference of the magnetic structures between our previous model. Hence, to develop theoretical considerations, the infinite-range interaction is the important example of the long-range interaction as the first step of the consideration. The outline of this paper is as follows. We introduce the model and define its time development in Sec.~\ref{Model}, investigate the thermodynamic limit and finite size system in Secs.~\ref{MF} and \ref{Simulation}, respectively, and summarize the study in Sec.~\ref{Summary}.

\section{Model \label{Model} }

To consider the effect of the interaction range, we introduce the present model, which resembles that of our previous study~\cite{Komatsu19}, but differs in having an infinite-range interaction. The model in this study is composed of the two Ising spin lattices adjacent to each other, as in our previous model. We also let the magnetic order of this model behave as the potential barrier, which prevents the relative motion of lattices like the antiferromagnetic order of our previous model does.
In the short-range model of magnetism, antiferromagnetic order is composed of two sublattices that have oppositely directed magnetization. To consider the infinite-range model without destroying this structure, we first divide each lattice into two sublattices and assume that the coupling constant between spins depends only on the sublattices to which they belong, not on the distance between them, as we explain next.
We name the sublattices in the upper lattice $U_A$ and $U_B$, and those in the lower lattice $D_A$ and $D_B$. Each sublattice has $N/4$ lattice points with Ising spin variables $\sigma_i$. In each lattice, lattice points of two sublattices are arranged alternately, and the distance between adjacent lattice points is normalized as the unit length.
 We introduce the shift of the upper lattice, $r$, and let the pair of sublattices, $U_A$ and $D_A$ (or $U_B$ and $D_B$), be closest to each other when $r=0$. When $0<r<1$, $U_A$ ($U_B$) goes away from $D_A$ ($D_B$) and approaches $D_B$ ($D_A$) with increasing $r$, and when $1<r<2$, the opposite occurs. The positional relation between sublattices reverts to that of $r=0$ when $r=2$, then repeats the pattern of $0 \leq r <2$ periodically when $r > 2$. We do not need other geometrical restrictions in this model because the infinite-range interaction does not decay by the distance between spins. 
 
 We let the coupling constant between the spins of the same sublattice be $\frac{2J'}{N}$, that between the spins of different sublattices of the same lattice (namely, $U_A$ and $U_B$ or $D_A$ and $D_B$) be $\frac{J}{N}$, and that between the spins of different lattices be the periodic function of $r$, $\frac{J  h_1(r)}{N}$ or $\frac{J h_2(r)}{N}$. Here, we let $h_1(r)$ and $h_2 (r)$ be the periodic and piecewise linear function, which takes its maximum value, 1, when the two sublattices are closest, and its minimum value, 0, when they are farthest. That is, 

%\begin{eqnarray}
\begin{subequations}
\begin{align}
h_1 (r) & =  \left\{
\begin{array}{c}
1-r \ \ \ \mathrm{if} \ \ \ 0 \leq r < 1, \\
r-1 \ \ \ \mathrm{if} \ \ \ 1 \leq r < 2, \\
\end{array} 
\right. \label{h1_def} \\
h_2 (r) & = \left\{
\begin{array}{c}
r \ \ \ \mathrm{if} \ \ \ 0 \leq r < 1, \\
2-r \ \ \ \mathrm{if} \ \ \ 1 \leq r < 2, \\
\end{array} 
\right. \label{h2_def}
\end{align}
\end{subequations}
%\end{eqnarray}
The periods of $h_1$ and $h_2$ are given as 2: $ h_1 (r+2) = h_1 (r) $, $h_2 (r+2) = h_2 (r)$. We can also consider the model with different forms of $h_1$ and $h_2$. However, to discuss the effect of the interaction range by comparing this model with the short-range model in our previous study, Eqs. (\ref{h1_def}) and (\ref{h2_def}), which coincides with the periodic extension of the intersurface interaction of the previous model except for the constant $\frac{1}{N}$, is the appropriate form.
Under this coupling constant, sublattices of different lattices tend to be closer if their spins are oriented oppositely, and farther apart if their spins are oriented in the same direction. This effect acts as a potential barrier, which prevents lattice motion. Sublattices with different directions have an important role in preventing the lattice motion in our model. Considering this point, it is difficult to develop a similar discussion in the ferromagnetic model where uniform magnetization appears. According to previous studies, such as Ref. \cite{KHW08}, it is fluctuations of the spins that prevent the lattice motion of the ferromagnetic models. However, in the case of the infinite-range interaction, even the contribution of the fluctuation is small because it is averaged over the whole system. This is why we use the antiferromagnetic model even though the structure with sublattices is complicated.

\begin{figure}[hbp!]
\begin{center}
\includegraphics[width = 7.0cm]{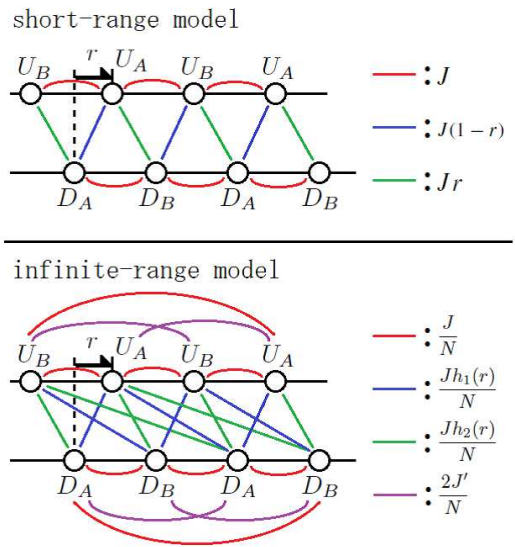}
\end{center}
\caption{(Color online) Comparison of the short-range model in our previous study \cite{Komatsu19} (top) and the infinite-range model in this study (bottom). In the top panel, the red, blue, and green lines indicate the pairs of spins with coupling constants of $J$, $J(1-r)$, and $Jr$, respectively. In bottom panel, the red, blue, green, and purple lines indicate the pairs of spins with coupling constants of $\frac{J}{N}$, $\frac{J  h_1(r)}{N}$, $\frac{J  h_1(r)}{N}$, and $\frac{2J'}{N}$. }
\label{Lattice1}
\end{figure}

The difference between the short-range model of our previous study and this model is shown in Fig. \ref{Lattice1}. The relation between different sublattices in this model resembles that of the short-range model, even though the interaction range is lengthened. There is no interaction between the spins of same sublattice (purple lines in Fig. \ref{Lattice1}) in the short-range model. This interaction is added to generalize the discussion.
 
The Hamiltonian of this system is given as 
\begin{widetext}
\begin{eqnarray}
{\cal H} & = & \frac{J}{N} \left( \sum _{ i_A \in U_A, i_B \in U_B } \sigma _{i_A} \sigma _{i_B} + \sum _{ j_A \in D_A, j_B \in D_B } \sigma _{j_A} \sigma _{j_B} \right) \nonumber \\
& & + \frac{J h_1(r)}{N} \left( \sum _{ i_A \in U_A, j_A \in D_A } \sigma _{i_A} \sigma _{j_A} + \sum _{ i_B \in U_B, j_B \in D_B } \sigma _{i_B} \sigma _{j_B} \right) \nonumber \\
& & +  \frac{J h_2(r)}{N} \left( \sum _{ i_A \in U_A, j_B \in D_B } \sigma _{i_A} \sigma _{j_B} + \sum _{ i_B \in U_B, j_A \in D_A } \sigma _{i_B} \sigma _{j_A} \right) \nonumber \\
& & + \frac{J'}{N} \left( \sum _{ i_A , i' _A \in U_A } \sigma _{i_A} \sigma _{i' _A} + \sum _{ i_B , i' _B \in U_B } \sigma _{i_B} \sigma _{i' _B} + \sum _{ j_A , j' _A \in D_A } \sigma _{j_A} \sigma _{j'_A} + \sum _{ j_B , j' _B \in D_B } \sigma _{j_B} \sigma _{j'_B} \right) .
\label{Hamiltonian}
\end{eqnarray}
\end{widetext}
Here, the coupling constant between the spins of the same sublattice is halved considering the double counting in the summation.

We let the magnetizations of the sublattices per one spin be $m_{U_A}$, $m_{U_B}$, $m_{D_A}$, and $m_{D_B}$. Namely, $m_{U_A} \equiv \left( \frac{N}{4} \right) ^{-1} \sum _{i_A \in U_A} \sigma _{i_A} $, for example. Using these parameters, Eq. (\ref{Hamiltonian}) is simplified as
\begin{eqnarray}
{\cal H} 
 & = & \frac{NJ}{16} \left\{ \left( m_{U_A} m_{U_B} + m _{D_A} m _{D_B} \right) \right. \nonumber \\
 & & + h_1 (r) \left( m_{U_A} m_{D_A} + m _{U_B} m _{D_B} \right) \nonumber \\
 & & + \left. h_2 (r) \left( m_{U_A} m_{D_B} + m_{U_B} m_{D_A} \right) \right\} \nonumber \\
 & & + \frac{NJ'}{16} \left( m_{U_A} ^2 + m_{U_B} ^2 + m _{D_A}^2 + m _{D_B}^2 \right).  \label{Hamiltonian2}
\end{eqnarray}
Note that the number of spins of each sublattice is $\frac{N}{4}$, and the magnetizations defined above are not the ensemble averages.
 As in our previous study~\cite{Komatsu19}, we define the time development of this model by updating of the Metropolis method and let the unit of time be 1 Monte Carlo step (MCS). The proposed updating at each step is the reversal of one randomly chosen spin. Note that this dynamics changes only one spin at each step of updating. Hence, we cannot impose the symmetry stemming from the antiferromagnetic order, i.e. $m_{U_A} = - m_{U_B}$ or $m_{D_A} = - m_{D_B}$, on the Hamiltonian itself, because this imposition requires the complicated updating of the system which changes more than two spins simultaneously. 

We fix the lower lattice and consider the motion of the upper lattice by imposing an external force, $F$, on it. This force balances the frictional force in the steady state. The shift of the upper lattice, $r$, obeys the overdamped Langevin equation under given temperature $T$. In this work, we let the Boltzmann constant $k_B =1$ by adjusting the unit of temperature. Assuming that any elastic deformation of the lattices can be ignored, the Langevin equation is written as  
\begin{equation}
0 = - \gamma \left( \frac{N}{2} \right) \frac{dr}{dt} + F - \frac{\partial {\cal H} }{\partial r} + \sqrt{2 \gamma T \left( \frac{N}{2} \right) } R (t),
\label{Langevin0}
\end{equation}
where $R$ is the white Gaussian noise fulfilling $\left< R(t) R(t') \right> = \delta (t-t')$. Using the external force per lattice point $f \equiv 2F/N $, we transform Eq.~(\ref{Langevin0}) as
\begin{equation}
\frac{dr}{dt} = \frac{f}{\gamma} + \frac{2}{\gamma N} \left( - \frac{\partial {\cal H} }{\partial r} + \sqrt{\gamma T N} R (t) \right).
\label{Langevin1}
\end{equation}

\section{Thermodynamic limit \label{MF} }
The mean field approximation describes the behaviors of infinite-range models well, and it becomes exact at the thermodynamic limit. Hence, in this section, we consider this model by using the mean field approximation. In the mean field approximation, fluctuations of thermodynamic quantities are ignored and these quantities nearly coincide with the ensemble averages of themselves, for example, $\left< m_{U_A} \right> \simeq m_{U_A} $. Hence, we do not distinguish the expressions of these two quantities.
 
\begin{figure}[hbp!]
\begin{center}
\begin{minipage}{0.99\hsize}
\includegraphics[width = 8.0cm]{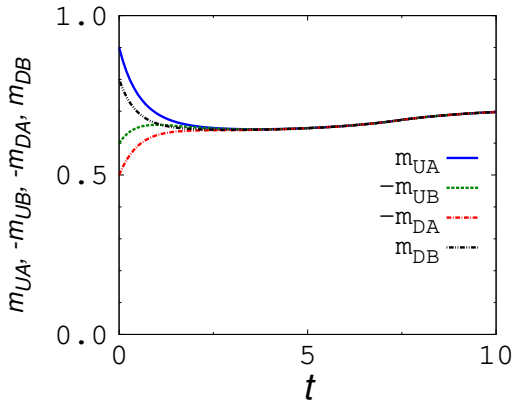} 
\end{minipage}
\end{center}
\caption{(Color online) An example of the time-dependence of parameters $m_{U_A}$(blue solid line), $-m_{U_B} $(green dashed line), $-m_{D_A}$(red dash-dotted line), and $m_{D_B} $(black dash-double-dotted line) obtained by the numerical calculation of Eqs. (\ref{Langevin1}) and (\ref{dmUB}) using the fourth-order Runge--Kutta method. In this example, we let $T=0.4$, $f=0.2$, $J=1$, $J'=0$, and $\gamma = 1$, and the initial state is given as $m_{U_A} = 0.9$, $m_{U_B} = -0.6$, $m_{D_A} = -0.5$, $m_{D_B} = 0.8$, and $r=0.1$. To confirm that the solution converges to the symmetric one obeying Eq. (\ref{m_assumption}), we plot $-m_{U_B} $ and $-m_{D_A} $, instead of $m_{U_B} $ and $m_{D_A} $ themselves. }
\label{m_example}
\end{figure}

The probability that one up (down) spin of sublattice $U_A$ is reversed at each step of updating, $P_{\mathrm{u \rightarrow d} } $($P_{\mathrm{d \rightarrow u} } $), is given as 
\begin{subequations}
\begin{align}
P_{\mathrm{u \rightarrow d} } & = \frac{1}{4} \left( \frac{1+m_{U_A} }{2} \right) \min \left[ 1, \exp \left( -\beta \delta {\cal H} _{\mathrm{u \rightarrow d} } \right) \right], \label{Pud} \\
P_{\mathrm{d \rightarrow u} } & = \frac{1}{4} \left( \frac{1-m_{U_A} }{2} \right) \min \left[ 1, \exp \left( -\beta \delta {\cal H} _{\mathrm{d \rightarrow u} } \right) \right]. \label{Pdu}    
\end{align}
\end{subequations}
In these equations, $\frac{1}{4} \frac{1 \pm m_{U_A} }{2}$ is the probability that each up or down spin is chosen as the candidate for updating, and the minima are the acceptance rates of the proposal.
The change in $M_{U_A}$ when an up or down spin is reversed is $\mp 2$, so the corresponding change in $m_{U_A}$ is given as 
\begin{subequations}
\begin{align}
\delta m_{\mathrm{u \rightarrow d} } & = - 2 \cdot \left( \frac{N}{4} \right) ^{-1} = - \frac{8}{N} \label{mud} \\
\delta m_{\mathrm{d \rightarrow u} } & = - \delta m_{\mathrm{u \rightarrow d} }. \label{mdu} 
\end{align}
\end{subequations}
Hence, the change in the Hamiltonian caused by each update is given as
\begin{subequations}
\begin{align}
\delta {\cal H} _{\mathrm{u \rightarrow d} } & = \frac{NJ}{16} \left( -\frac{8}{N} \right) \left( m_{U_B} + h_1 (r) m_{D_A} + h_2 (r) m_{D_B} \right) \nonumber \\ & \ \ \ \  + \frac{NJ'}{16} \left( -\frac{8}{N} \right) \cdot (2 m_{U_A})  \nonumber \\
& =  -\frac{J}{2} \left( m_{U_B} + h_1 (r) m_{D_A} + h_2 (r) m_{D_B} \right) \nonumber \\ & \ \ \ \ - J' m_{U_A}, \label{Hud} \\
\delta {\cal H} _{\mathrm{d \rightarrow u} } & =  -\delta {\cal H} _{\mathrm{u \rightarrow d} }. \label{Hdu} 
\end{align}
\end{subequations}
The change in the Hamiltonian at each step is described by the mean value of the magnetization, $m_{U_A}, m_{U_B}, m_{D_A}$, and $m_{D_B}$, because we use the infinite-range model described by Eq. (\ref{Hamiltonian2}).
Using Eqs. (\ref{mud}) and (\ref{mdu}), the total change of $m_{U_A}$ per time interval $\delta t = \frac{1}{N}$, namely the one-step updating, is expressed as
\begin{eqnarray} 
\delta m_{U_A} & = & P_{\mathrm{u \rightarrow d} } \delta m_{\mathrm{u \rightarrow d} } + P_{\mathrm{d \rightarrow u} } \delta m_{\mathrm{d \rightarrow u} } \nonumber \\
& = & \frac{8}{N} (-P_{\mathrm{u \rightarrow d} } + P_{\mathrm{d \rightarrow u} } ). 
\end{eqnarray}
From this relation, the time development of $m_{U_A}$ is given as
\begin{equation}
\frac{d m_{U_A} }{dt} \simeq \frac{\delta m_{U_A} }{\delta t} = 8 (-P_{\mathrm{u \rightarrow d} } + P_{\mathrm{d \rightarrow u} } ).
\label{dm1}
\end{equation}
Substituting Eqs. (\ref{Pud}), (\ref{Pdu}), (\ref{Hud}), and (\ref{Hdu}) into Eq. (\ref{dm1}), we finally get
\begin{widetext}
\begin{eqnarray}
\frac{d m_{U_A} }{dt} & = & - (1 + m_{U_A} ) \min \left[ 1, \exp \left\{ \frac{\beta J}{2} \left( m_{U_B} + h_1 (r) m_{D_A} + h_2 (r) m_{D_B} \right) + \beta J' m_{U_A} \right\} \right] \nonumber \\
& & + (1 - m_{U_A} ) \min \left[ 1, \exp \left\{ - \frac{\beta J}{2} \left( m_{U_B} + h_1 (r) m_{D_A} + h_2 (r) m_{D_B} \right) - \beta J' m_{U_A} \right\} \right].
\label{dmUA}
\end{eqnarray}
\end{widetext}
The time development equations of $m_{U_B}, m_{D_A}$, and $m_{D_B}$ are derived in a similar way,
\begin{widetext}
\begin{eqnarray}
\frac{d m_{S_{\mu} } }{dt} & = & - (1 + m_{S_\mu} ) \min \left[ 1, \exp \left\{ \frac{\beta J}{2} \left( m_{ S_{\bar{\mu}} } + h_1 (r) m_{ \bar{S}_{\mu} } + h_2 (r) m_{ \bar{S}_{\bar{\mu}} } \right) + \beta J' m_{S_{\mu} } \right\} \right] \nonumber \\
& & + (1 - m_{S_{\mu} } ) \min \left[ 1, \exp \left\{ - \frac{\beta J}{2} \left( m_{S_{\bar{\mu}} } + h_1 (r) m_{\bar{S}_{\mu} } + h_2 (r) m_{\bar{S}_{\bar{\mu}} } \right) - \beta J' m_{S_{\mu} } \right\} \right] \label{dmUB}. 
\end{eqnarray}
\end{widetext}
Here, $S=U,D$, $\bar{S} =D,U$, $\mu =A,B$, and $\bar{\mu} = B,A$. Note that $S \neq \bar{S}$, $\mu \neq\bar{\mu}$ in every case.
We assume that the two sublattices of each lattice have the opposite magnetization, that is, $m_{U_A} = - m_{U_B} $ and $m_{D_A} = - m_{D_B} $. Furthermore, from the symmetry between the upper and the lower lattice, we can infer that $m_{U_A}$ is equal to $m_{D_A}$ or $m_{D_B}$. Choosing the latter case, we obtain the relation
\begin{equation}
m_{U_A} = - m_{U_B} = - m_{D_A} = m_{D_B} \equiv m.
\label{m_assumption}
\end{equation}
To confirm that Eq. (\ref{m_assumption}) is valid, we investigated some initial conditions which slightly break the symmetry of this equation by the numerical calculation of Eqs. (\ref{Langevin1}) and (\ref{dmUB}), and found that the difference from the symmetric solution always attenuated like the example shown in Fig. \ref{m_example}.

We should also refer to the possibility that the antiferromagnetic order itself does not appear. Generally speaking, even in the antiferromagnetic model, it is possible that the lattice motion introduces the effective ferromagnetic interaction and the ferromagnetic order appears, i.e. $m_{U_A} = m_{U_B} $ and $ m_{D_A} =m_{D_B} $. However, in the case of this model, the interactions between different lattices are so weak that the energy of the ferromagnetic state cannot be stabler than that of the antiferromagnetic one at any $r$.

Using Eq. (\ref{m_assumption}), Eq. (\ref{dmUB}) is simplified as
\begin{widetext}
\begin{eqnarray}
\frac{d m }{dt} & = & - (1 + m ) \min \left[ 1, \exp \left\{ \frac{\beta }{2} \left( \left( -1 - h_1 (r) + h_2 (r) \right) J + 2J' \right) m  \right\} \right] \nonumber \\
& & + (1 - m ) \min \left[ 1, \exp \left\{ - \frac{\beta }{2} \left( \left( -1 - h_1 (r) + h_2 (r) \right) J + 2J' \right) m \right\} \right] \label{dmU}.
\end{eqnarray}
\end{widetext}

The time development of $r$ is described by Eq. (\ref{Langevin1}). Considering that the random force term of this equation disappears at the thermodynamic limit, this equation is expressed as
\begin{eqnarray}
\frac{dr}{dt} & = & \frac{f}{\gamma} - \frac{2}{\gamma N}  \frac{\partial {\cal H} }{\partial r} \nonumber \\
& = & \frac{f}{\gamma} + \frac{J}{4 \gamma} \left( h_1 ' (r) - h_2 ' (r) \right) m^2.
\label{Langevin2}
\end{eqnarray} 
Using Eqs. (\ref{dmU}) and (\ref{Langevin2}), we can discuss the time development of the whole system at the thermodynamic limit.
The trivial solution of these equations is the case where the magnetizations do not exist,
\begin{equation}
 m = 0, \ \ \frac{dr}{dt} = \frac{f}{\gamma}.
\label{solution_para}
\end{equation}
In this case, the upper lattice moves with constant velocity $f/\gamma$, so the system obeys the Stokes law. 

One more important solution is the case where the magnetizations have a nonzero constant value, namely, $m \neq 0$, and $\frac{d m }{dt} = 0$. In this case, Eq. (\ref{dmU}) is transformed as
%\begin{widetext}
%\begin{eqnarray}
%0 & = & - (1 + m_{U} ) \min \left[ 1, \exp \left\{ \frac{\beta J}{2} \left( -m_{U} + (h_1 (r) - h_2 (r) ) m_{D} \right) \right\} \right] \nonumber \\
%& + & (1 - m_{U} ) \min \left[ 1, \exp \left\{ - \frac{\beta J}{2} \left( -m_{U} + (h_1 (r) - h_2 (r) ) m_{D} \right) \right\} \right] \label{dmU2}. \\
%\end{eqnarray}
%\end{widetext}
%Transforming this equation, we obtain 
\begin{widetext}
\begin{eqnarray}
\frac{1+m}{1-m} & = & \frac{ \min \left[ 1, \exp \left\{ - \frac{\beta }{2}  \left( \left( -1 - h_1 (r) + h_2 (r) \right) J + 2J' \right) m \right\} \right] }{ \min \left[ 1, \exp \left\{ \frac{\beta }{2}  \left( \left( -1 - h_1 (r) + h_2 (r) \right) J + 2J' \right) m \right\} \right] } \nonumber \\
& = & \exp \left\{ - \frac{\beta }{2} \left( \left( -1 - h_1 (r) + h_2 (r) \right) J + 2J' \right) m \right\}. \label{dmU3} 
\end{eqnarray}
\end{widetext}
Using this equation, we get
\begin{equation}
m =  \tanh \left\{ - \frac{\beta }{4} \left( \left( -1 - h_1 (r) + h_2 (r) \right) J + 2J' \right) m \right\}. \label{dm3t} 
\end{equation}
Eq. (\ref{dm3t}) is the self-consistent equation that determines the values of magnetization $m$. From these equations, $r$ should also be constant to keep the magnetizations constant. However, Eq. (\ref{Langevin2}) usually does not have a solution that simultaneously fulfills $\frac{dr}{dt} = 0$ and $m \neq 0$. This seems to be a contradiction, but we can avoid this problem if $r$ has the value near the discontinuous point of $ h_1 ' (r) $ and $ h_2 ' (r) $, that is, $r \simeq 0$ or 1. This result means that the upper lattice is perfectly trapped by the potential barrier created by the magnetic interaction. First, we consider the case of $r \simeq 0$. In this case, the term $h_1 ' (r) - h_2 ' (r) $ that appears in Eq. (\ref{Langevin2}) has the value -2 or +2 when $r$ is slightly larger or smaller than 0, so Eq. (\ref{Langevin2}) can be transformed as
\begin{equation}
\frac{dr}{dt} = \frac{f}{\gamma} - \frac{J}{2 \gamma} m ^2 \mathrm{sgn} (r).
\label{Langevin3}
\end{equation} 
To keep the value of $r$ near 0, the right-hand side of Eq. (\ref{Langevin3}) should be negative when $r>0$ and positive when $r<0$. This condition is satisfied when and only when 
\begin{equation}
|f| < \frac{J}{2} m ^2, \label{fstop}
\end{equation}
so we call this upper limit $f_c$, that is, $f_c \equiv J m ^2 / 2$. 
In the case where $r \simeq 1$, we can transform Eq. (\ref{Langevin2}) as
\begin{equation}
\frac{dr}{dt} = \frac{f}{\gamma} + \frac{J}{2 \gamma} m ^2 \mathrm{sgn} (r-1)
\label{Langevin3a}
\end{equation} 
by a similar calculation. Eq. (\ref{Langevin3a}) shows that $r=1$ is the unstable point because the right-hand side of this equation is always positive when $r>1$. Hence, $r \simeq 0$ is the only case we should consider.
Substituting $r \simeq 0$ into Eq. (\ref{dm3t}), we get
\begin{equation}
m = \tanh \left( \frac{\beta }{2} (J-J') m \right). \label{dm5}  
\end{equation}
This self-consistent equation has the nonzero solution only when 
\begin{equation}
\frac{\beta }{2} (J-J') > 1.  
\end{equation}
This inequation is equivalent to $T < T_c \equiv (J-J')/2$. If the temperature is higher than the critical value, $T_c$, the system cannot have nonzero magnetization and $f_c = 0$. In particular, in the case where $J'>J$, nonzero magnetization never appears because $T_c < 0$. 

\begin{figure}[hbp!]
\begin{center}
\begin{minipage}{0.99\hsize}
\includegraphics[width = 8.0cm]{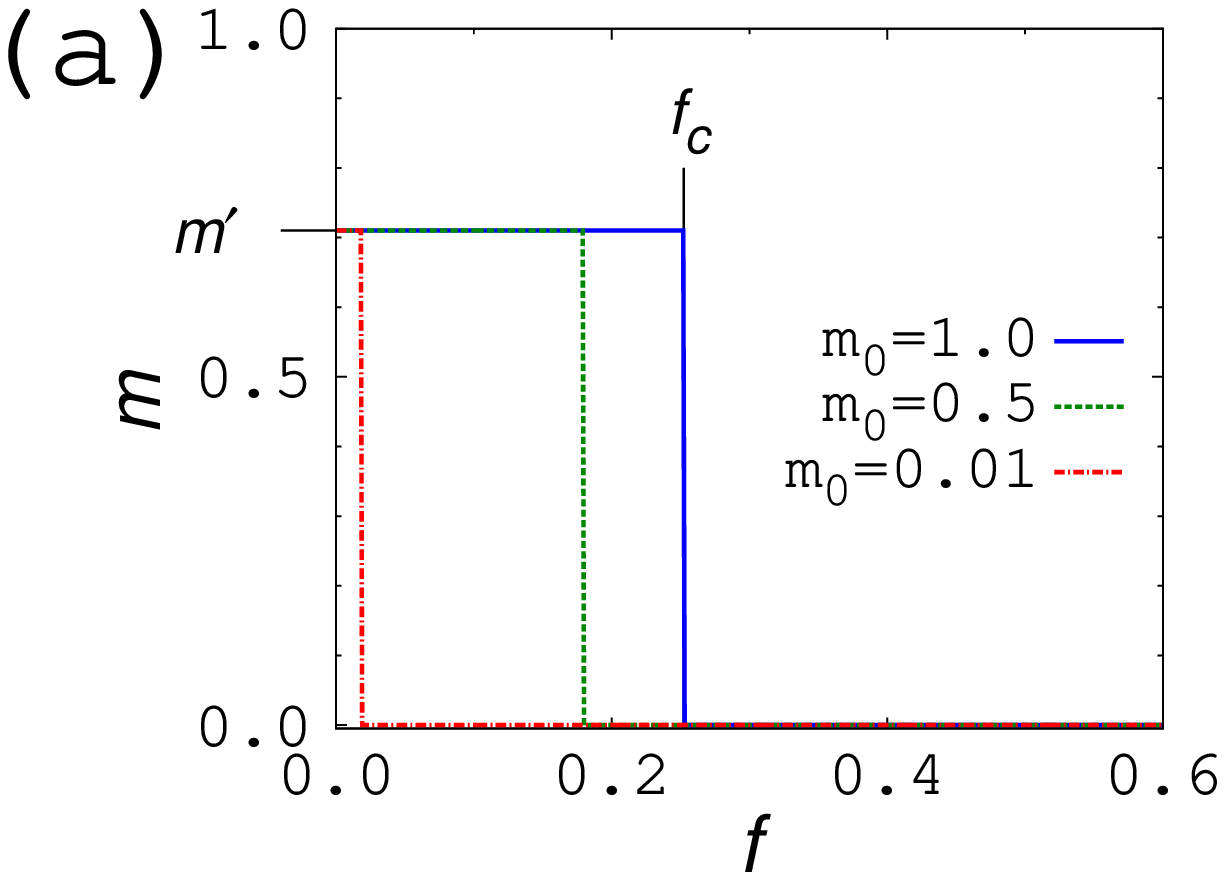} \\
\includegraphics[width = 8.0cm]{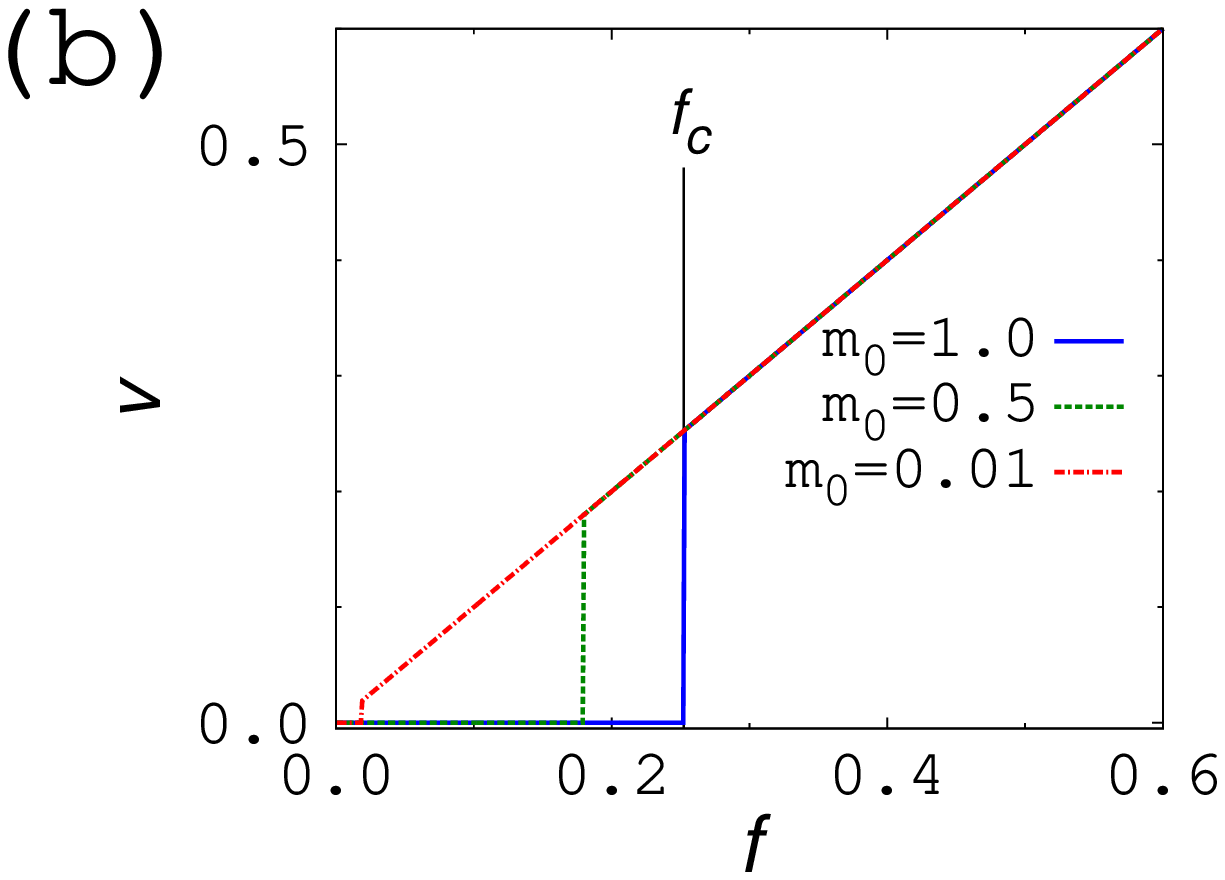} 
\end{minipage}
\end{center}
\caption{(Color online) $f$-dependence of (a) magnetization $m$ and (b) velocity $v$ at $T=0.4(=0.8T_c)$ at the thermodynamic limit. The blue solid, green dashed, and red dash-dotted lines represent cases where $m_0 = 1.0, 0.5, 0.01$, respectively. In each case, $m$ is equal to $m '$, the nontrivial solution of the self-consistent equation (\ref{dm5}), when $f$ is small. The value of $f_c$ is defined after Eq. (\ref{fstop}). In the case of $T=0.4$, these values are given as $m' \simeq 0.710$, and $f_c \simeq 0.252$ }
\label{mvRK04}
\end{figure}

\begin{figure}[hbp!]
\begin{center}
\begin{minipage}{0.99\hsize}
\includegraphics[width = 8.0cm]{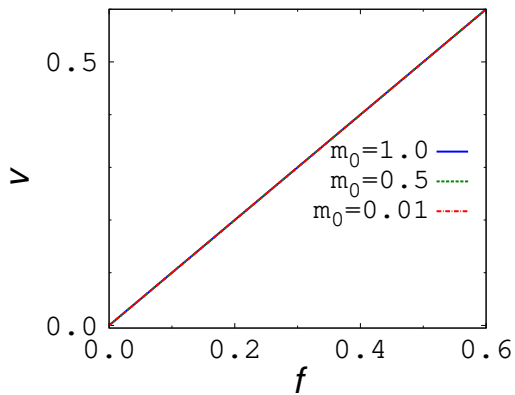} 
\end{minipage}
\end{center}
\caption{(Color online) $f$-dependence of velocity $v$ at $T=0.5(=T_c)$ at the thermodynamic limit. The meanings of the lines are the same as in Fig. \ref{mvRK04}. In this case, the system obeys the Stokes law. The graph for $m$ is not plotted because it is always zero. }
\label{mvRK05}
\end{figure}

We investigate the actual dependence of $m$ and velocity $v \equiv \frac{dr}{dt}$ on $f$ in the steady state of the system that obeys Eqs. (\ref{dmU}) and (\ref{Langevin2}) by the fourth-order Runge--Kutta method. In this calculation, we let $J=1, J'=0$, and $\gamma=1$, and start from the initial state $m = m_0, r=0$. Here, several cases with different $m_0$ are investigated. Each measured quantity is averaged over $1.0 \times 10^4 \leq t \leq 5.0 \times 10^4$. The result is shown in Figs. \ref{mvRK04} and \ref{mvRK05}. 
 Here, the graph of $m$ at $T=0.5$ is not plotted because $m$  is always zero at this temperature. Seeing these graphs, two types of states are observed. In one, magnetization $m$ coincides with $m '$, the nontrivial solution of the self-consistent equation (\ref{dm5}), and velocity $v$ is zero. In the other, $m = 0$ and $v$ obey the Stokes law; $v = f/\gamma$. Hence, there are no other steady state solutions of Eqs. (\ref{dmU}) and (\ref{Langevin2}) than the two types discussed above. When the temperature is lower than the critical value, $T_c$, both of these states are observed, and the value of $f$ where they interchange depends on the initial value of $m_U$ within the range of $f \leq f_c$. In contrast, when $T \geq T_c$, the system always obeys the Stokes law and does not have the magnetization. It is notable that the hysteresis dependence is observed in the low-temperature state.
 
 The discussion of this section is valid only when $N = \infty$. In a finite size system, the magnetizations do not coincide with the ensemble averages of themselves. Namely, $\left< m_{U_A} \right> \neq m_{U_A} $, for example. The fluctuations of the magnetizations, which is the difference between their real values and ensemble averages, is $O(1/\sqrt{N})$. Therefore, the effect of this fluctuation is comparable to the random force term of the Langevin equation (\ref{Langevin1}), which we could ignore in this section. This fact makes the analytical discussion of the finite size system difficult, so a numerical simulation is necessary.

\section{Simulation \label{Simulation}}

\begin{figure}[hbp!]
\begin{center}
\begin{minipage}{0.99\hsize}
\includegraphics[width = 8.0cm]{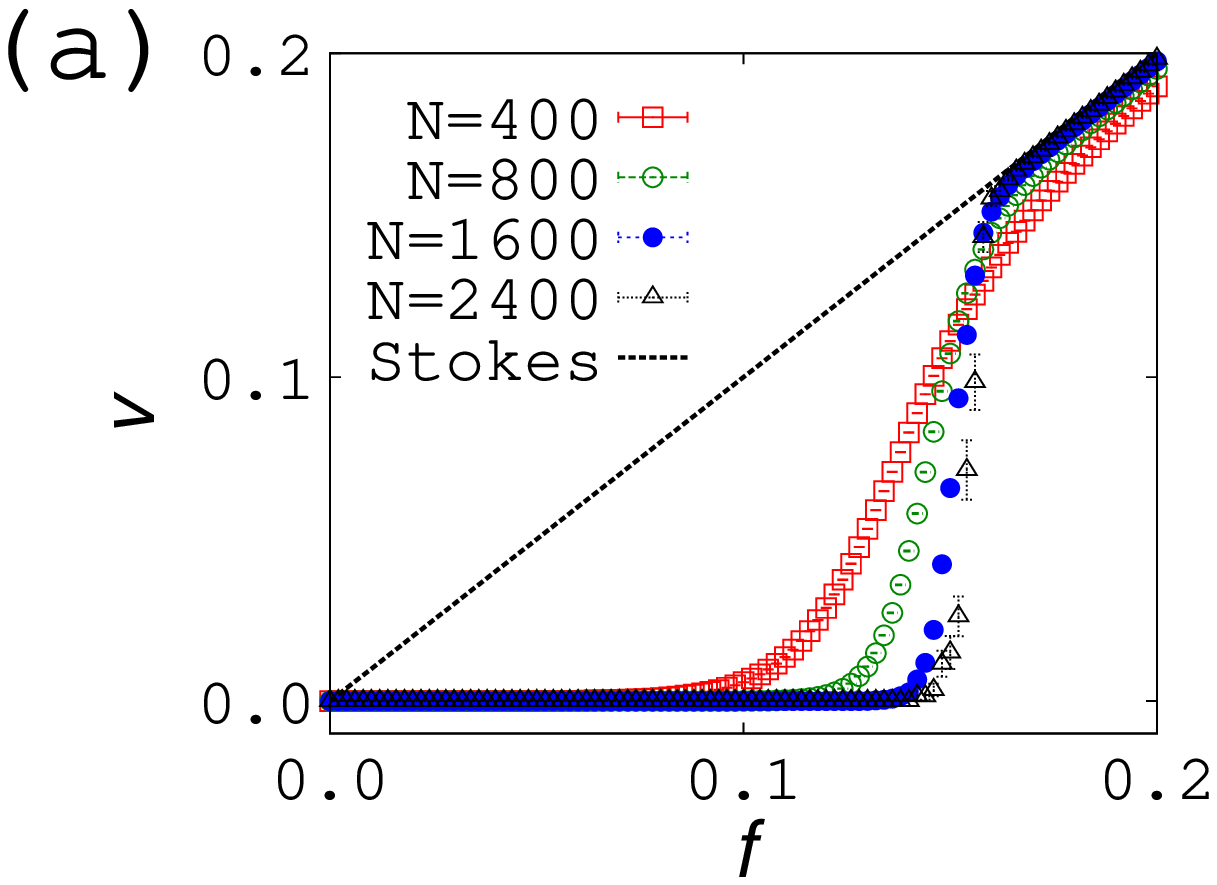} \\ 
\includegraphics[width = 8.0cm]{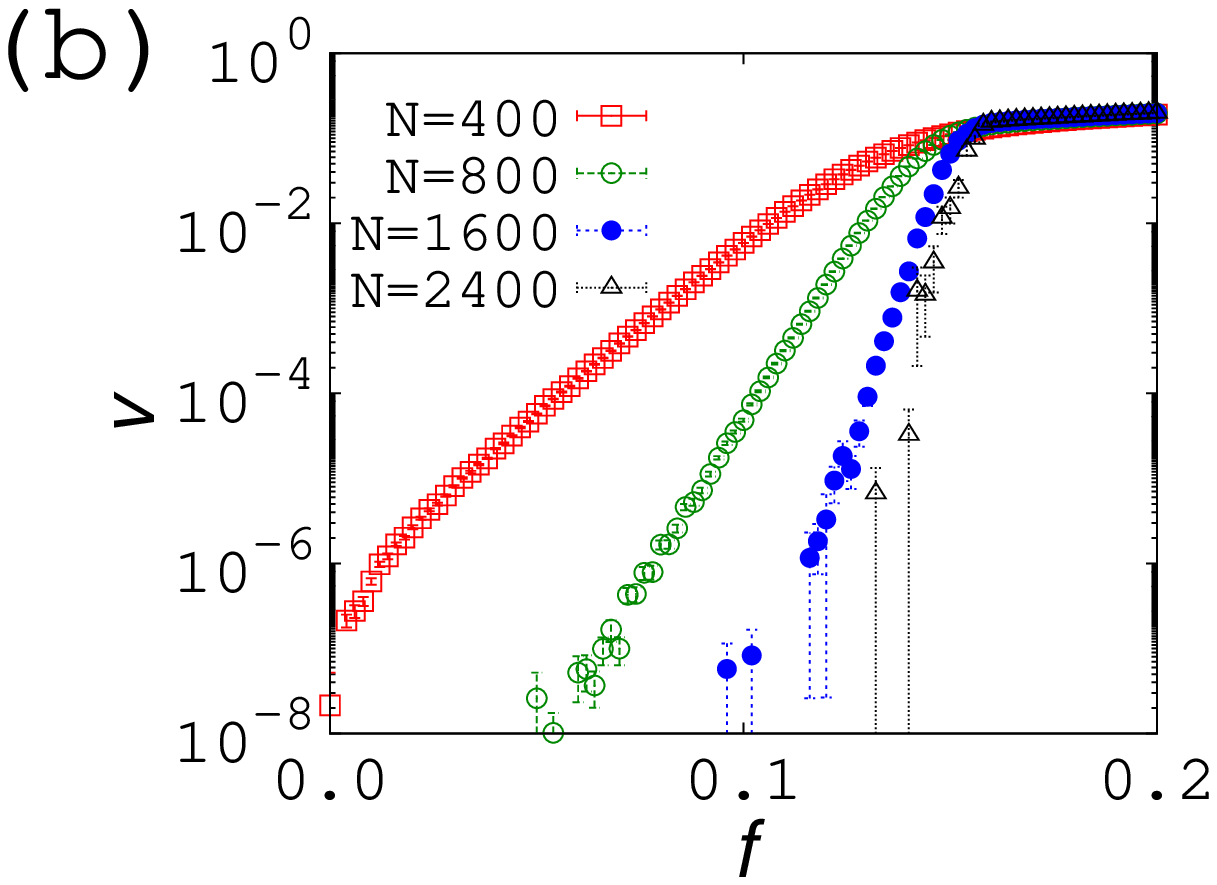}  \\
\includegraphics[width = 8.0cm]{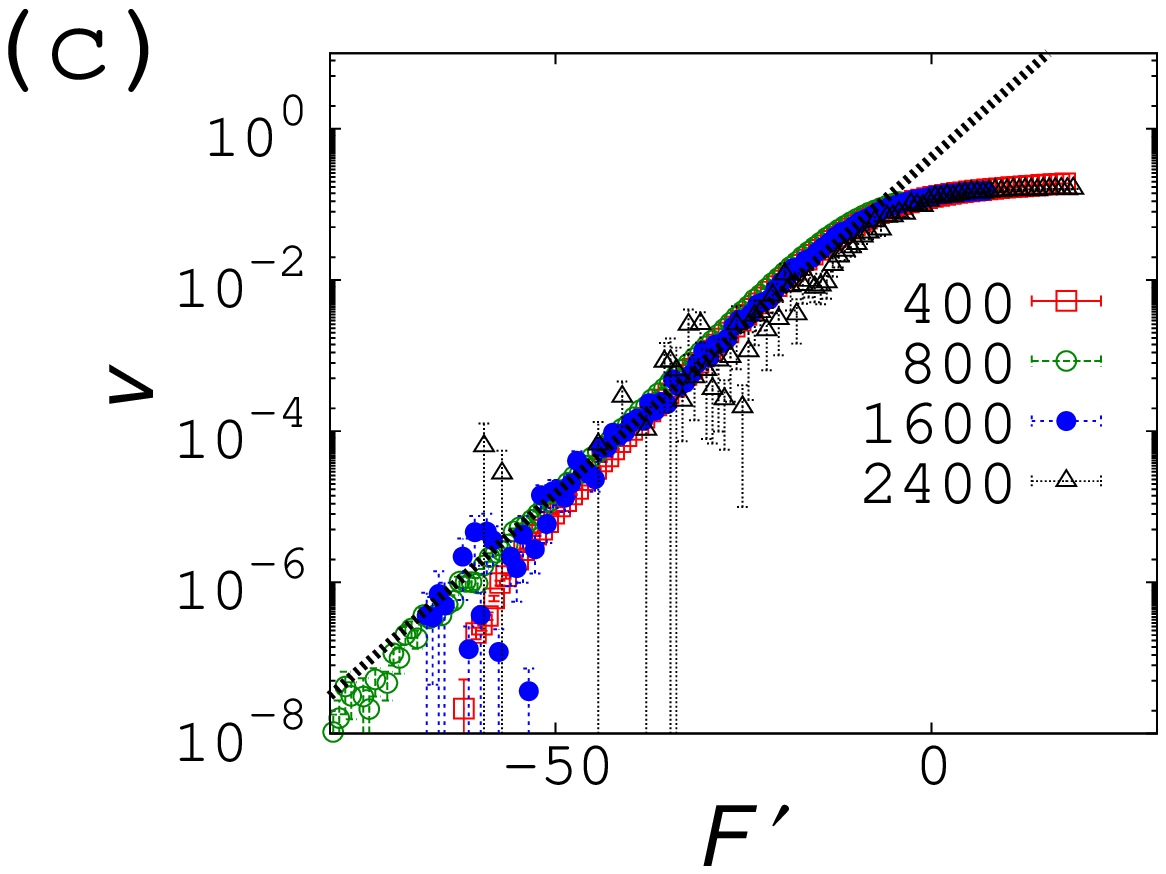} 
%\hspace{4.0cm}
\end{minipage}
\end{center}
\caption{(Color online) $v$--$f$ relation at $T=0.4$ plotted on a (a) linear graph and (b) semi-logarithmic graph. The red open squares, green open circles, blue closed circles, and black open triangles indicate data at $N=400$, 800, 1600, and 2400, respectively, and the black dotted line of the graph (a) represents Stokes law $v = f/\gamma$. (c) Rescaled $v$--$F'$ curves at $T=0.4$. The symbols are the same as in (a) and (b), and the black dotted line is the fitting curve. The fitting parameters are given as $a = 0.204 \pm 0.004$, $f_c ' = 0.155 \pm 0.004$, and $c = 0.88 \pm 0.13$.  }
\label{fv04}
\end{figure}

In this section, we investigate the behavior of this model by a numerical simulation. $r$ is updated after every $\Delta t$ MCSs ($=N \Delta t$ steps) by applying the stochastic Heun method to Eq.~(\ref{Langevin1}).
We let $J=1, J'=0, \gamma = 1$ and $\Delta t = 0.01$; thus, $r$ is updated after every $N /100$ steps of updating the spin variables. Velocity $v$ of the upper lattice is defined as the change in $r$ per MCS.
We begin the simulation from the perfectly ordered state and increase $f$ gradually. At each value of $f$, the first $1.0 \times 10^6 $ MCSs are used for relaxation and the next $4.0 \times 10^6$ MCSs are used for measurement, and after that, $f$ is increased by $\Delta f = 0.002$. The initial value of $r$ is given as 0. We first investigate the $v$--$f$ curve at $T=0.4 (=0.8T_c)$, where $T_c$ is the critical value discussed in the previous section. At each calculation, we take the average over 48 independent trials to obtain the data with error bars. The result is given as Figs.~\ref{fv04} (a) and (b). Based on these graphs, the system shows a crossover or transition from the Dieterich--Ruina law to the Stokes law. 
To consider the mechanism explaining why the Dieterich--Ruina law appears, we rescale these $v$--$f$ curves with a relation similar to that in our previous study~\cite{Komatsu19}. The equation we used for the rescaling in our previous study was derived by applying the discussion in Ref.~\cite{HBPCC94} to the magnetic friction model. This discussion is based on the assumption that $v$ is determined mainly by the competition between external force and the potential barrier. In this study, the equation is expressed as
\begin{equation}
\log v = a F' + c,
\label{DRlaw2}
\end{equation}
\begin{equation}
\mathrm{where} \ \ F' = N \left( f- f_c ' \right). \label{Fscale}
\end{equation}
The change from the corresponding equation in our previous study is that the parameter expressing the contact area between surfaces is replaced by $N$. This is because the height of the potential barrier is proportional to the contact area in our previous model, and to the total number of spins, $N$, in the present model.
The rescaled curves are plotted in Fig.~\ref{fv04} (c). Here, we change the value of $\Delta f$, which is the gradual increase of $f$ after measurement, into $\Delta f = 0.8/N$. Namely, $\Delta f$ is made inversely proportional to $N$ to take sufficient data even when $N$ is large. The fitting parameters, $a, f_c ', c$, are determined by the least-squares fitting, in which we use data points that satisfy $10^{-6} \leq v \leq 10^{-2}$. The graph at $N=2400$ has a larger error than the other graphs. Hence, we impose the same weight at every point, regardless of the error bar, to fit this graph well. The rescaled graphs overlap with the fitting curve, so the mechanism explaining why this model obeys the Dieterich--Ruina law resembles that of our previous study.

\begin{figure}[hbp!]
\begin{center}
\begin{minipage}{0.99\hsize}
\includegraphics[width = 8.0cm]{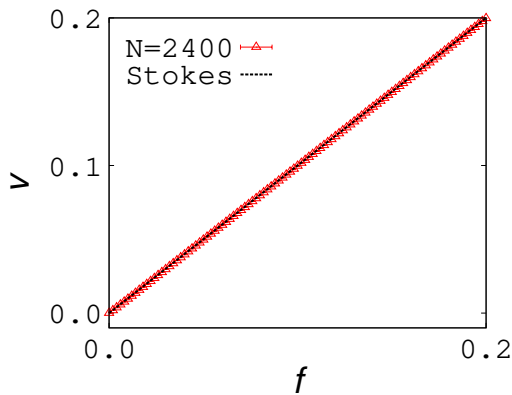} 
\end{minipage}
\end{center}
\caption{(Color online) $v$--$f$ relation at $T=0.5$ and $N=2400$. The data indicated by the red open triangles coincide with the Stokes law represented by the black dotted line. }
\label{fv05}
\end{figure}

We also investigate the $v$--$f$ relation at $T=0.5(=T_c)$, the temperature at which the self-consistent equation (\ref{dm5}) does not have the nonzero solution. The result is plotted in Fig.~\ref{fv05}. We investigate the $v$--$f$ relation of $N \leq 1600$ at this temperature, but we show the plot for only $N=2400$ because the shape of the $v$--$f$ curve is almost unchanged by $N$. This graph shows that the system always obeys the Stokes law. This behavior is similar to that of the case of the thermodynamic limit discussed in the previous section, but quite different from that of the short-range model of our previous study~\cite{Komatsu19}, which shows a crossover or transition from the Dieterich--Ruina law to the Stokes law regardless of the temperature. This difference results from the point that the frictional force of the present model depends directly on the long-range order and disappears when the order becomes zero.

Comparing this section and the previous one, the results of the simulation and the mean field analysis are qualitatively different from each other if $T < T_c$. In the case of the simulation of the finite size system, the $v$--$f$ curve is divided into two domains, the domain where the lattice motion is trapped by the magnetic structure and that where the magnetic order disappears and the lattice can move. The value $f _c '$ evaluated by the finite size scaling of this section is the approximate value of the border between these domains. %In the case of $T=0.4$, for example, this value is given as $f_c ' \simeq 0.155 $.
The result of the mean field analysis, on the other hand, the system shows the hysteresis dependence. Namely, if $f$ is smaller than a certain value $f _c$ given in the previous section, the system has the two types of steady states corresponding to the two domains of finite size system, and which one of them appears depends on the initial condition. Note that these two quantities $f_c '$ and $f_c $ have the different values from each other. In the case of $T=0.4$, for example, $f_c ' \simeq 0.155$ and $f_c \simeq 0.252$.(See the captions of Figs. \ref{mvRK04} and \ref{fv04}.)
 
 As we have pointed out in the previous section, the mean field analysis is thought to become exact in the thermodynamic limit $N \rightarrow \infty$, so the difference between these two investigations seems to be a contradiction. To consider why this difference occur and see how the result of simulation approaches that of the mean field analysis in $N \rightarrow \infty$, we calculate the time-dependence of the order parameter for several system sizes and conditions. In the actual simulation, we prepare the three cases, (a), (b), and (c). At each calculation, we take the average over 4800 independent trials to obtain the data with error bars.
In the case (a), $f$ is given as $f = 0.1(<f_c ')$, and the initial condition is set as $m_{UA} = -m_{UB} = -m_{DA} = m_{DB} = 0.1$, and $r=0$. In the case (b) and (c), $f$ is given as $f=0.2$(which fulfills $f_c ' < f < f_c$) and $f=0.3(>f_c)$, respectively. Initial conditions of these two cases are the perfectly ordered states $m_{UA} = -m_{UB} = -m_{DA} = m_{DB} = 1$, with $r=0$.
 We also calculate the same value under the same conditions by the mean field analysis using Eqs. (\ref{dmU}) and (\ref{Langevin2}), and compare the results with those of the simulation.

\begin{figure}[!hbp]
\begin{center}
\includegraphics[width = 8.0cm]{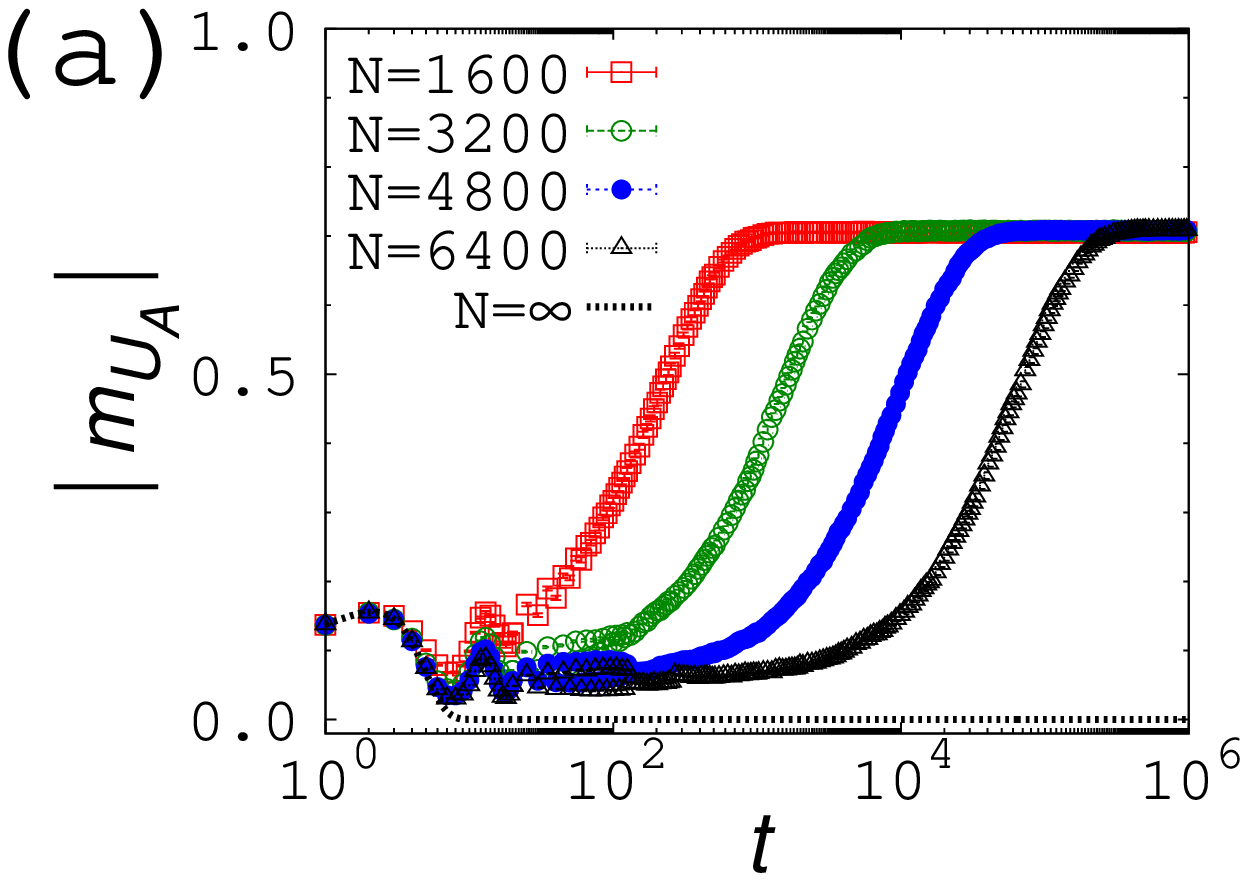} \\
\includegraphics[width = 8.0cm]{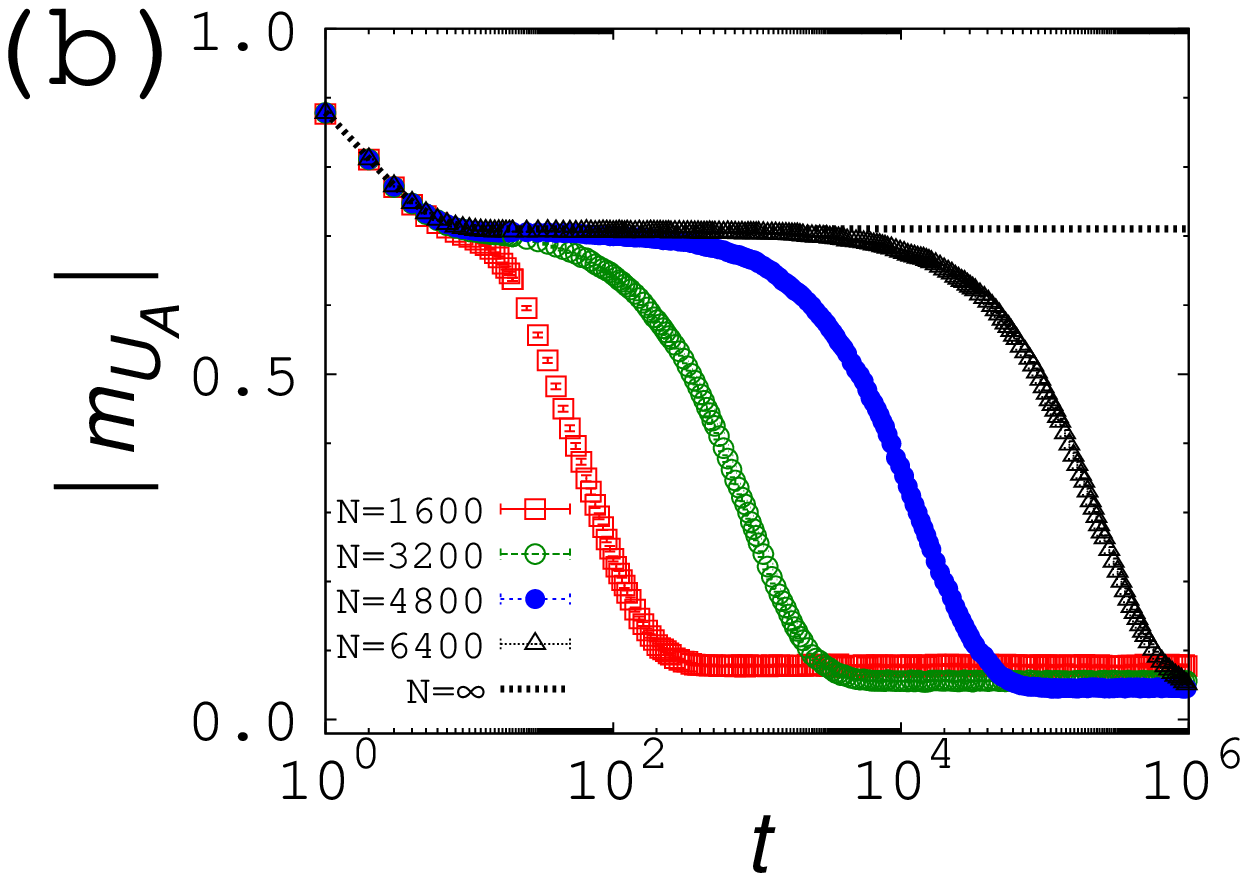} \\
\includegraphics[width = 8.0cm]{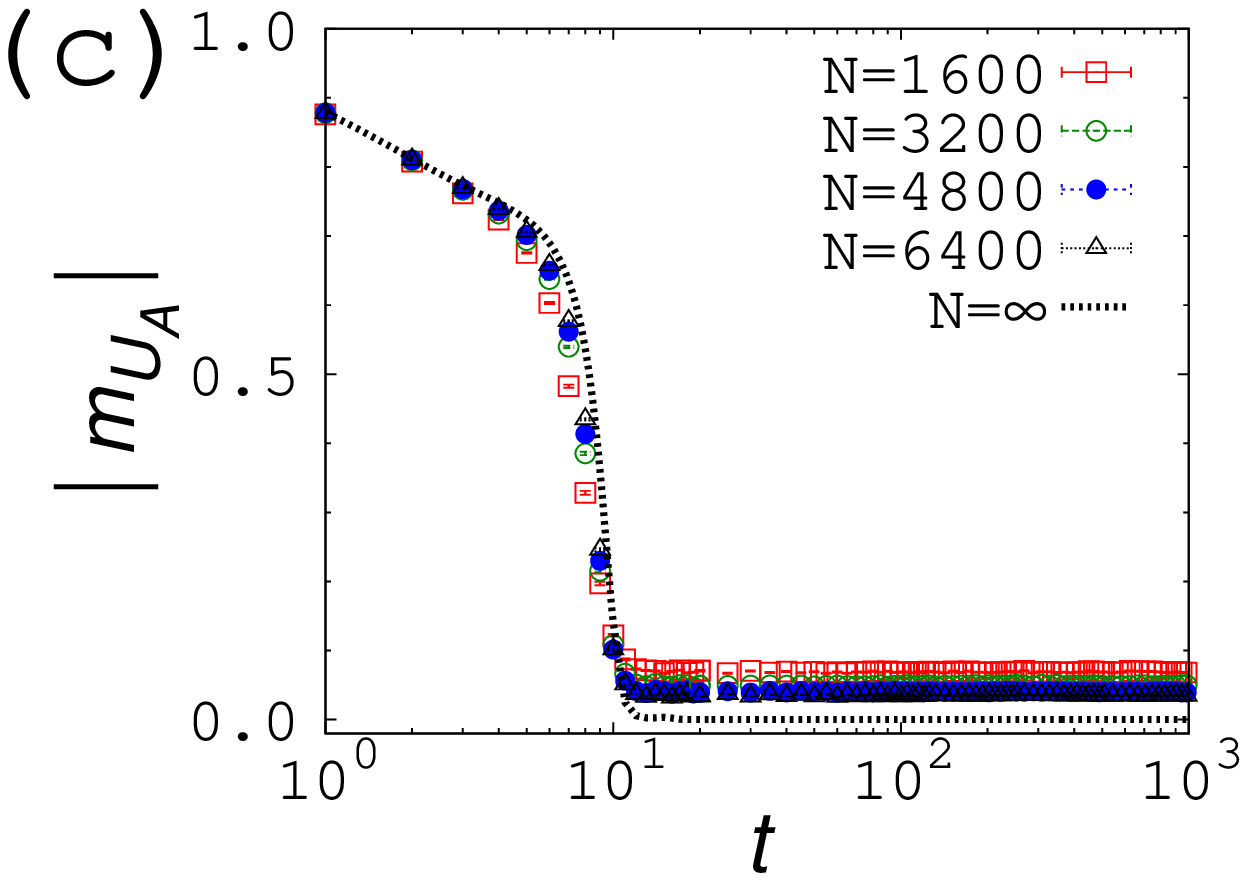} \\
\caption{(Color online) Time dependence of $m_{UA}$ in three cases, (a).$f=0.1$, (b).$f=0.2$, and (c).$f=0.3$. The initial condition for each case is explained in the main text. The red open squares, green open circles, blue closed circles, and black open triangles indicate the simulation data at $N=1600$, 3200, 4800, and 6400, respectively, and the black dotted lines are the result of the mean field analysis using Eqs. (\ref{dmU}) and (\ref{Langevin2}).  }
\label{time_dep}
\end{center}
\end{figure}

The results are shown in Fig. \ref{time_dep}. Judging from Figs. \ref{time_dep}.(a) and (b), relaxation of the system slows down with increasing system size $N$ when $f < f_c$. In the thermodynamic limit $N \rightarrow \infty$, the relaxation time diverges, and the new steady state which is not observed in the finite size system appears. This is why the result of mean field analysis has the hysteresis dependence. The symptom of this new steady state is observed as the plateaus of finite $N$ graphs of Figs. \ref{time_dep}.(a) and (b). Note that the reason why $m_{UA}$ has the slightly different value from zero in the plateau of Fig. \ref{time_dep}.(a) is the effect of the $O(1/\sqrt{N})$ fluctuation. In the case of $f>f_c$, on the other hand, according to Figs. \ref{time_dep}. (c), the time-dependence of the system hardly depends on $N$. 

From these graphs, we can confirm that the qualitative difference between finite and infinite size systems is caused by the divergence of the relaxation time, and two values $f_c '$, the border of the domains in the finite size system, and $f_c$, the upper limit of $f$ at which the hysteresis dependence exists in the thermodynamic limit, can be defined without contradiction.  

\section{Summary \label{Summary}}

We introduced a model of magnetic friction and investigated its behavior, including the $v$--$f$ relation. This model resembles our previous model, but has an infinite-range interaction instead of a short-range interaction. We used two methods for different situations, namely, the mean field analysis for the thermodynamic limit and the numerical simulation for the finite size system. 
\begin{table}[htb]
  \begin{tabular}{c|c|c|c}
  \hline \hline
   & short-range & \multicolumn{2}{c}{infinite-range} \\
  & $N < \infty$ & $N < \infty$ & $N \rightarrow \infty$ \\ \hline
  $T<T_c$ & D-R to Stokes & D-R to Stokes & hysteresis \\ \hline
  $T \geq T_c$ & D-R to Stokes & Stokes & Stokes \\
    \hline \hline
  \end{tabular}
  \caption{$v$--$f$ relation of the short-range model in Ref.~\cite{Komatsu19} and the infinite-range model in this study. ``D-R to Stokes'' means that the model shows a crossover or transition from the Dieterich-Ruina law to the Stokes law, ``Stokes'' means that the model always obeys the Stokes law, and ``hysteresis'' means that the model has hysteresis dependence. For the short-range model, $T_c$ is the equilibrium transition temperature. }
  \label{T_vf}
\end{table}

The $v$--$f$ relation in this model is summarized in Table~\ref{T_vf}. Note that we do not know the behavior at the thermodynamic limit, $N \rightarrow \infty$, for the short-range model because we cannot derive the exact result of this case.  Hence, we cannot determine whether the existence of the hysteresis dependence observed in this study is related to the interaction range. The main difference between this model and the short-range model in our previous study is the temperature dependence of the behavior. When the temperature is higher than the critical value, $T_c$, the present model always obeys the Stokes law, whereas the short-range model shows a crossover or transition from the Dieterich--Ruina law to the Stokes law even when the temperature is higher than the equilibrium transition temperature. This is because the frictional force of the present model depends directly on the long-range order and disappears when the order becomes zero. When the temperature is lower than $T_c$, the model shows a crossover or transition similar to that of the short-range model when $N < \infty$, whereas it shows hysteresis dependence when $N \rightarrow \infty$. According to the last discussion of section \ref{Simulation}, divergence of the relaxation time at $N \rightarrow \infty$ is the cause of this hysteresis dependence. The difference between the finite and infinite size systems results from the random force term of the Langevin equation and the fluctuations of the sublattice magnetizations. To consider the finite size system, we cannot ignore either the random force term or these fluctuations because they make the contributions comparable to each other, namely, $O(1/\sqrt{N})$. 

Studies such as Ref. \cite{Hucht09} point out that the sufficiently fast motion of the lattice lets the short-range interaction system behave like a mean field system, so the short-range model may have the dynamical transition temperature we could not find in our previous study and show the qualitatively same behavior as the infinite-range model. However, considering that the lattice motion in our model is slow when the lattice is trapped by the potential barrier made by the magnetic structure, it is unclear whether this inference is correct\footnote{ Strictly speaking, in the case  of the model considered in Ref. \cite{Hucht09}, the critical value of the velocity, which divide the mean field-like and non-mean field-like behaviors, shows the power law decay with increasing the contact area at least when the spatial dimension $d=2$. It means that this model shows the mean field-like behavior at any finite velocity in the thermodynamic limit. This fact seems to be contradictory to the discussion of section \ref{Summary}. However, in the cases of our models of this paper and Ref. \cite{Komatsu19}, the velocity in the low-$f$ domain shows exponential decay, the faster decay than the critical value of the model of Ref. \cite{Hucht09}, with increasing the contact area.(See Eqs. (\ref{DRlaw2}) and (\ref{Fscale}) of this paper or Eq. (11) of Ref. \cite{Komatsu19}.) Hence, the velocity of these cases is small compared with the upper limit of the non-mean field-like domain of Ref. \cite{Hucht09}, if the system size is sufficiently large. }.
It is also unclear whether the behavior of the systems with realistic long-range interactions, such as the dipolar interaction, resembles the infinite-range interaction system or the short-range system. These problems should be investigated in future work.

\section*{Acknowledgments}
Part of the numerical calculations was performed on the Numerical Materials Simulator at the National Institute for Materials Science.

%\clearpage

\end{document}